**Visible Light-Activated Photosensitization of Hybridization of Far-red Fluorescent Protein and Silk**

*Jung Woo Leem, Jongwoo Park, Seong-Wan Kim, Seong-Ryul Kim, Seung Ho Choi, Kwang-Ho Choi, and Young L. Kim*

Dr. J. W. Leem, Dr. S. H. Choi, Prof. Y. L. Kim
Weldon School of Biomedical Engineering
Purdue University
West Lafayette, IN 47907, USA

Dr. J. Park, Dr. S. W. Kim, Dr. S. R. Kim, Dr. K. H. Choi
Department of Agricultural Biology
National Institute of Agricultural Sciences
Rural Development Administration
Wanju, Jeollabuk-do 55365, Republic of Korea

Prof. Y. L. Kim
Regenstrief Center for Healthcare Engineering
Purdue University
West Lafayette, IN 47907, USA

Prof. Y. L. Kim
Purdue Quantum Center
Purdue University
West Lafayette, IN 47907, USA

Abstract: Fluorescent proteins often result in phototoxicity and cytotoxicity, in particular because some red fluorescent proteins (RFP) produce and release reactive oxygen species (ROS). The photogeneration of ROS is considered as a detrimental side effect in cellular imaging or is proactively utilized for ablating cancerous tissue. As ancient textiles or biomaterials, silk produced by silkworms can directly be used as fabrics or be processed into materials and structures to host other functional nanomaterials. We report that transgenic fusion of far-red fluorescent protein (mKate2) with silk provides a photosensitizer hybridization platform for photoinducible control of ROS. Taking advantage of green (visible) light activation, native and regenerated mKate2 silk can produce and release superoxide and singlet oxygen, in a comparable manner of visible light-driven plasmonic photocatalysis. Thus, the genetic expression of mKate2 in silk offers immediately exploitable and scalable photocatalyst-like biomaterials. We further envision that mKate2 silk could potentially rule out hazardous concerns associated with foreign semiconductor photocatalytic nanomaterials.

Visible light-driven plasmonic photocatalysis, which relies on the combination of semiconductor photocatalysts with metal nanostructures/nanoparticles, has received considerable attention for solar energy conversion and utilization.[1] Solar photocatalysis has a variety of energy and environmental applications, such as hydrogen generation, carbon dioxide reduction, desalination, disinfection, and water/air purification.[1c,2] Specifically, the production of reactive oxygen species (ROS) photoinduced from photocatalyis has direct utilization for environment remediation and biomedicine. However, such applications are often intrinsically limited for large-scale and mass production. In addition, potentially hazardous and adverse (e.g. carcinogenic and cytotoxic) effects associated with semiconductor nanoparticles have limited the widespread utilization.[3] In this respect, we take inspiration from nature to identify and characterize plasmonic photocatalyst-like biological materials and further translate them into industrially relevant production processes.

The phototoxicity of fluorescent proteins, in particular red fluorescent proteins (RFP) is unanimously acknowledged in several different scientific communities; Some of RFP generate and release ROS upon light excitation, while the exact types of ROS vary among different RFP variants.[4] Since the use of such RFP was restricted by cytotoxicity, noncytotoxic RFP variants have been successfully developed for whole-cell labeling and cellular imaging *in vivo*.[5] In contrast, cytotoxic RFP has also been used as a means of selectively damaging specific proteins upon light activation, which is known as chromophore-assisted light inactivation (CALI).[6] In the latter case, RFP is recapitulated as 'genetically encoded ROS-generating proteins' for inactivating target cells and ablating tissue of interest.[4d] All of these characteristics of RFP suggest that semiconductor nanocrystals or conjugated nanoparticles for plasmonic photocatalysis can be replaced by phototoxic RFP.

Some fluorescent proteins participate in Type I and Type II photosensitization reactions.[7] Predominant ROS generated by fluorescent proteins depends on the type of photosensitization reactions and the concentration of local molecular oxygen (i.e. electron acceptor). For example, (enhanced) green fluorescent protein, (E)GFP typically produces singlet oxygen ($^1O_2$) via Type II photosensitization reaction, in which energy transfer occurs from the excited triplet state of the fluorescent protein to molecular oxygen.[8] RFP, such as KillerRed, can undergo Type I photosensitization reaction, in which electron transfer to molecular oxygen yields superoxide ($O_2^{\bullet-}$).[6c,9] Another interesting aspect of ROS resulting from Type I and Type II photosensitization reactions is that the maximum migration (or damage) distance of $O_2^{\bullet-}$ and $^1O_2$ is less than 200 – 300 nm, depedening on the surrounding envirменents.[6a,b,10,11] This relatively short damage distance can be advantagous as a safeguard, given that $O_2^{\bullet-}$ and $^1O_2$ are instantaneously reactive and toxic. Importantly, the resultant ROS (i.e. $O_2^{\bullet-}$ and $^1O_2$) generated by plasmonic photocatalysis using visible light is the same as that of RFP photosensitization reactions.[6c,9,12]

In this work, we introduce biological hybridization of far-red fluorescent proteins and natural proteins (i.e. silk) for a new class of genetically encoded photosensitization that can be activated by visible (or solar) light, producing selective ROS in a similar manner of plasmonic photocatalysis. Direct detection of ROS is known to be highly challenging, because ROS is extremely reactive and unstable. Thus, we implement several different approaches using turn-on/off fluorescent radical probes and physical quenchers/scavengers to experimentally validate ROS generated by transgenic RFP silk upon green light activation. We demonstrate that transgenic RFP silk can be mass-produced by scalable and continuous manufacturing using the currently available textile infrastructure. Using the polymeric nature of silk, transgenic RFP silk is further processed into nanomaterials and nanostructures in a variety of forms. The use of plasmonic photocatalyst-like proteins can overcome the limitation of potential adverse effects associated with foreign synthesized nanoparticles. We also envision that this bioreactor approach could potentially offer an alternative green manufacturing strategy for next generation photocatalysts.

We provide the impetus of visible light-activated genetically encoded ROS-generating multifunctional biomaterials, by exploiting silk containing recombinant RFP produced by transgenic silkworms (*Bombyx mori*) (**Figure 1**a). Silk produced by silkworms has extensively been utilized as fabrics and processed into engineered biomaterials due to its various merits of the superior mechanical and optical properties as well as the biocompatibility and biodegradability.[13] In particular, we take advantage of genetically engineered domesticated silkworms; transgenes of interests are expressed by germline transformation using the gene splicing method *piggyBac*.[13f,14] This silkworm transgenesis method can yield transformed animals with multiple successive generations and can produce recombinant substances in large amounts. The manufacturing processes of photocatalytic semiconductor nanoparticles often involve negative environmental consequences.[3] On the other hand, silkworm transgenesis enables us to readily produce natural photocatalysts and photosensitizers in an eco-friendly manner, minimizing the use of industrial facilities. Regarding an ecological hazard, it is highly unlikely that transgenic silkworms pose threats to natural ecosystems, because silkworms are dependent on humans for survival and reproduction as a completely domesticated indoor insect.

We choose mKate2, which is a far-red monomeric fluorescent protein.[5a] From a phototoxicity standpoint, mKate and mKate2 are widely considered as one of the cytotoxic standards.[5b-d] From a protein structural standpoint, the phototoxic action of mKate is commonly acknowledged to originate from a cleft-like opening channel filled with water molecules inside, allowing for enhanced generation and release of ROS. Specifically, mKate has a cleft-like $\beta$-barrel frame between $\beta$ sheets ($\beta 7$ and $\beta 10$), resulting in relatively high phototoxicity.[4e,15] Several other fluorescent proteins, including KillerRed,[16] SuperNova,[6c] KillerOrange,[17] Dronpa,[18] TurboGFP,[19] and mCherry,[20] have a similar $\beta$-barrel structure with a water-filled pore, which can also be used to tune the excitation wavelength range and to select the photosensitization properties. For the hybridization of mKate2 and silk, mKate2 gene is fused with N-terminal and C-terminal domains of the fibroin heavy chain promoter (pFibH); p3xP3-EGFP-pFibH-mKate2 is the constructed transformation vector (Figure 1b; Figures S1 and S2, Supporting Information). 3xP3-EGFP is served only for screening a large number of G1 broods, because EGFP fluorescent signals are easily monitored in the stemmata and the nervous system at early embryonic and larval stages. The silk gland of genetically encoded mKate2 silkworms is fluorescent (Figure 1c; Figure S3, Supporting Information).

The homogenous production of mKate2 silk results in a mass density of ~ 12.6% mKate2/Fibroin H-chain fusion recombinant protein.[14e] In Figure 1a, white (wild-type) silk cocoons are not fluorescent, while mKate2-expressing silk cocoons are fluorescent under green light excitation (Figure S4, Supporting Information).

In **Figure 2**a, we photometrically analyze the photocatalytic activity of mKate2 silk by degrading organic blue dye molecules (i.e. methylene blue) in an aqueous solution under green laser light activation ($\lambda_{ex}$ = 532 nm and optical intensity ≈ 0.2 mW mm$^{-2}$; Supporting Information) at the ambient room temperature. Although this crude method is not specific to particular types of ROS, the photodegradation of methylene blue serves as a standard for validating photocatalysis. However, silk has a strong affinity to organic molecules and metal ions.[14e,21] Thus, the loss of blue color in a methylene blue solution containing mKate2 silk is attributable to the infiltration (i.e. adsorption) of methylene blue to silk fibers as well as the photolysis of methylene blue itself by green light. In this respect, we performed separate degradation measurements to account for the adsorption of methylene blue to silk under a dark condition (i.e. no light irradiation) and the photolysis of methylene blue without any silk discs (Figure S5, Supporting Information). After factoring out these confounding effects, the contribution of ROS generated by the mKate2 silk is significant; a linear fit between $\ln(C_t/C_0)$ of methylene blue by mKate2 silk and the irradiation time $t$ results in an apparent pseudo-first-order rate constant ($k_{app}$) value of 2.46×10$^{-4}$ min$^{-1}$ (Inset of Figure 2a).

As a model system of testing ROS production, we also examine the phototoxicity of mKate2 silk on *Escherichia coli* (*E. coli*) upon green light activation (Figure 2b). Historically, ROS generated by conventional photocatalysis has extensively been validated by demonstrating their antimicrobial activities.[22] After DH5α *E. coli* cells are attached on silk discs (Inset of Figure 2b), illumination from an easily accessible green light-emitting diode (LED) source ($\lambda_{ex}$ = 530 nm with a FWHM of 30 nm and optical intensity ≈ 0.02 mW mm$^{-2}$; Supporting Information), which is ~ 10 times weaker than that of the green laser source above, is irradiated on the surface of white silk and mKate2 silk for 30 – 60 minutes at the ambient room temperature. Such green light activation is not only accessible from sunlight, but also belongs to the peak solar radiation spectrum. Dark controls are also maintained without any light irradiation. Colony-forming unit (CFU) counts show a statistically significant difference only in bacterial inactivation between irradiated (Light ON) and unirradiated (Light OFF) mKate2 silk for 60 minutes (multiple comparison *p*-value = 0.031) (Tables S1 and S2, Supporting Information). The survival rate of *E. coli* from mKate2 silk under weak green light activation (Light ON) is reduced to 45%, compared with the corresponding dark controls (Light OFF). This result supports the idea of green light-activated genetically encoded photosensitization as an alternative ROS generation route, completely avoiding the use of photocatalytic semiconductor nanoparticles.

We further investigate specific types of ROS produced by mKate2 silk upon green light activation ($\lambda_{ex}$ = 532 nm and optical intensity ≈ 0.2 mW mm$^{-2}$) (**Figure 3**). First, we detect $O_2^{\bullet-}$ generated by mKate2 silk via primarily Type I photosensitization reaction. The generation and release of $O_2^{\bullet-}$ are monitored using fluorescent radical probes; 4-[(9-acridinecarbonyl)amino]-2,2,6,6-tetramethylpiperidin-1-oxyl (TEMPO-9-ac) is commonly used to sense $O_2^{\bullet-}$.[9] Under consistent green light irradiation on mKate2 silk discs immersed in TEMPO-9-ac solutions, fluorescent signals of TEMPO-9-ac ($\lambda_{ex}$ ≈ 360 nm and $\lambda_{em}$ ≈ 440 nm) are detected in two different configurations (Figure S6, Supporting Information): i) The turn-on fluorescent radical probes on the mKate2 silk surface are diffused in the TEMPO-9-ac solution. In Figure 3a, the fluorescent emission intensity of TEMPO-9-ac increases monotonously with the duration of green light irradiation, compared the baseline signals before light activation (controls). ii) After TEMPO-9-ac is permeated into the silk discs, the turn-on fluorescent radical probes remain inside, which in turn emit blue fluorescence of TEMPO-9-ac from the mKate2 silk discs. 240-minute irradiation of green light leads to a 2-fold increase in the radical probe fluorescent intensity from the silk discs infiltrated with TEMPO-9-ac, compared with the unirradiated mKate2 silk discs (controls) (Figure S7, Supporting Information). These results are in excellent agreement with $O_2^{\bullet-}$ released from KillerRed, which is one of the highly phototoxic RFP variants.[9] Second, we detect $^1O_2$ generated by mKate2 silk via Type II photosensitization reaction under the same green light activation, using 9,10-anthracenediyl-bis(methylene)dimalonic acid (ABDA) as a radical probe. While the original state of ABDA emits fluorescence ($\lambda_{ex} \approx 380$ nm and $\lambda_{em} \approx 431$ nm), ABDA reacts with $^1O_2$ to yield endoperoxide as a turn-off fluorescent radical probe, reducing its fluorescent intensity.[23] In Figure 3b, the intensity of ABDA fluorescent peaks gradually drops as the irradiation time increases, supporting the generation of $^1O_2$.

Using fluorogenic scavengers, we additionally validate the generation of $O_2^{\bullet-}$ and $^1O_2$ from Type I and Type II photosensitization reactions of mKate2 silk. The phototoxicity of RFP is always accompanied by photobleaching, because the formation of ROS itself facilitates the degradation of RFP excitation-emission cycles.[24] Interestingly, TEMPO-9-ac and ABDA, which are fluorescent radical probes, can also be used as physical quenchers of $O_2^{\bullet-}$ and $^1O_2$, respectively, without directly reacting with other free radicals (Figure S8, Supporting Information).[25] In Figures 3c and 3d, the inhibition of TEMPO-9-ac and ABDA in photobleaching of mKate2 silk provides another level of evidence, supporting $O_2^{\bullet-}$ and $^1O_2$ production. In other words, the uptake of local surrounding ROS ($O_2^{\bullet-}$ and $^1O_2$) prevents mKate2 silk from being photodamaged, which is manifested by the relatively sustained fluorescent intensity of mKate2 silk. In a mixed solution of TEMPO-9-ac and ABDA, the fluorescent emission of mKate2 silk is further maintained (Figure 3e). We also confirm reduced photobleaching of mKate2 silk using other scavengers of $O_2^{•-}$ (nitro blue tetrazolium chloride, NBT) and $^1O_2$ (sodium azide, $NaN_3$) (Figure S9, Supporting Information).[26]

The direct use of silk fibers produced by silkworms has its own advantage as utilized in the textile industry, because the transgenic silk has the comparable mechanical properties to wild-type silk to weave fabrics (Figure S10, Supporting Information). Silk fibroin can further be processed into polymeric materials for fabricating artificially engineered biomaterials and optical materials in a variety of forms with biocompatibility and bioabsorbablity.[13b-f] However, the conventional fibroin processing methods are inappropriate for mKate2 silk,[13c-e] because fluorescent proteins are highly susceptible to denaturation from high temperature and pH values.[13f,27] In our case, to minimize heat-induced denaturation of mKate2, mKate2 silk fibroin is extracted from silk cocoons at low temperature of 45 °C, assisted by alcalase enzyme and dithiothreitol (DTT) treatments.[13f] A reductase, such as DTT, is beneficial for renaturing the protein structure by reducing the disulfide bonds of proteins and peptides in a solvent.[28] In **Figures 4**a-4c, mKate2 silk fibroin is processed into an aqueous solution and then is formed into a flexible thin film. The fluorescent property of mKate2 is maintained in both of the regenerated mKate2 silk solution and the regenerated mKate2 silk film under green light excitation (Figure S11, Supporting Information). Importantly, the generation of $O_2^{•-}$ and $^1O_2$ from the regenerated mKate2 silk products is also detected using TEMPO-9-ac and ABDA, respectively (Figures 4d and 4e). With prolonged green light irradiation ($\lambda_{ex}$ = 532 nm and optical intensity ≈ 0.2 mW mm$^{-2}$), the fluorescent signal of TEMPO-9-ac increases, while that of ABDA decreases, supporting the two types of ROS generation. Similarly, the photodegradation of methylene blue by the mKate2 silk film results in $k_{app}$ = 1.12×10$^{-3}$ min$^{-1}$ under green light irradiation, after factoring out the confounding effects (i.e. adsorption and photolysis of methylene blue) (Figure S12, Supporting Information).

From an ecological standpoint, our results may suggest that the primary purpose of fluorescent proteins in nature could be photoinducible ROS generation, while the fluorescent emission may be a secondary consequence. From a mechanistic standpoint, ROS generation from fluorescent proteins is known to involve long-range electron transfer via two possible mechanisms of direct tunneling and hopping inside fluorescent proteins.[4c,29] The current understanding of this mechanism is based on quantum mechanics, because electron tunneling over such a long distance of 1.5 – 3 nm is typically impossible in vacuum.[30] From an electron donor standpoint, (E)GFP has been successfully tested for generating electricity as photodetectors and photovoltaics for bioenergy applications.[31] From a photocatalysis standpoint, the direct photosensitization properties of RFP have not yet been exploited for scalable photoreaction in a similar manner of plasmonic photocatalysis.

In conclusion, the reported hybridization of mKate2 and silk using genetically engineered silkworms can offer several pivotal advantages. Without a need of additional nanoconjugations (e.g. metals, dye molecules, and quantum dots), RFP can be excited by solar (visible) and green light, avoiding the most common carcinogen exposure of ultra-violet light. Both fluorescent proteins and silk are degradable and digestible,[32] eliminating the potential risk of exposure and consumption. As a biosynthesis reactor (i.e. green manufacturing), silkworm transgenesis is well-established for producing recombinant proteins in large amounts.[13f,14] As ancient textile materials, silk fibers are easily woven into large-area, continuous, and flexible fabrics using the existing textile manufacturing infrastructure.[14d,e] The unprecedentedly strong light scattering of native silk, which is manifested as the 'silvery' and 'lustrous' reflection,[33] can enhance interactions of light with RFP inside silk fibers.

**Experimental Section**

*Removal of sericin in silk (i.e. degumming)*: For effective generation and release of ROS from mKate2 silk, it was critical to remove the outermost layer (i.e. sericin) of silk fibers. We removed sericin using a degumming process. The outer sericin layer is commonly removed to improve the color, sheen, and texture of silk in the silk textile industry. However, conventional sericin removal methods are inappropriate for mKate2 silk, because these involve a boiling process in an aqueous solution.[13c-e] In our case, mKate2 silk cocoons were soaked in a pre-warmed mixture solution of sodium carbonate ($Na_2CO_3$, 0.2%) and Triton X100 (0.1%) at low temperature of < 60 °C under a vacuum pressure. During the degumming process, low pressure treatments (620 mmHg) were repeated several times to uniformly infiltrate the solution between silk fibers to remove most sericin. The degummed mKate2 silk cocoons were dried in dark under the ambient air conditions.

*Photodegradation of methylene blue as general photocatalytic quantification*: We quantified photodegradation of methylene blue, resulting from ROS generated by mKate2 silk under green light activation. For mKate2 silk specimens, silk cocoons were punched into 5-mm-diameter discs with a thickness of ~ 400 µm. We prepared methylene blue solutions (1 mL 0.05 wt.% methylene blue in 14 mL de-ionized water) containing 12 silk discs (total weight = 0.06 g). To reach the adsorption-desorption equilibrium in each test, the silk discs were stirred with 400 rpm in dark for two hours. Then, the silk discs were irradiated by green light ($\lambda_{ex}$ = 532 nm and optical intensity ≈ 0.2 mW mm$^{-2}$) for four hours, while being stirred. Aliquots (0.5 mL) were collected repeatedly with a fixed time interval and the spectral absorption of methylene blue was measured using a fiber bundle-coupled spectrometer with a white-light tungsten halogen source. To exactly quantify the photocatalytic activity of mKate2 silk only, separate degradation tests of methylene blue were also carried out to factor out two confounding effects: i) the adsorption of methylene blue to silk under a dark condition (i.e. no light irradiation) and ii) the photolysis of methylene blue without any silk discs by green light. For each elapsed irradiation time, a relative concentration $C_t/C_0$ of methylene blue was calculated using the absorption spectrum peak values $C_t$ at $\lambda = 668$ nm normalized by the absorption value $C_0$ before light irradiation (Figure 2a; Figure S5, Supporting Information). We estimated the reaction kinetics, following the apparent pseudo-first-order rate equation of Langmuir-Hinshelwood kinetics: $\ln(C_t/C_0) = - k_{app}t$, where $k_{app}$ is the rate constant (min$^{-1}$) and $t$ is the irradiation time (Insets of Figure 2a).

*Bacterial inactivation as general detection of ROS*: We tested ROS generated by mKate2 silk to inactivate *Escherichia coli* (*E. coli*) upon green light irradiation. We conducted four different groups of two different types of silk (i.e. white silk and mKate2 silk) and two light conditions (i.e. irradiation and unirradiation). We repeated these experiments for two different irradiation times of 30 and 60 minutes. Each bacterial inactivation experiment was performed in three assays with four replicates ($n = 12$) in each group for statistical analyses. DH5α *E. coli* cells were grown in a Luria-Bertani (LB) medium at 37 °C in a shaking incubator to an optical density at 600 nm (OD$_{600}$) of 2.5 ($\sim 2 \times 10^9$ cells mL$^{-1}$). The culture was diluted 10-fold and subsequently white silk and mKate2 silk discs (diameter = 6 mm) were placed on the culture. After incubation at 37 °C for 60 minutes, each silk disc was dried in dark for 30 minutes. For optical excitation of mKate2, the silk discs on a hydrated filter paper were irradiated with the green LED source ($\lambda_{ex} = 530$ nm with a FWHM of 30 nm and optical intensity $\approx 0.02$ mW mm$^{-2}$) for 30 – 60 minutes at the ambient room temperature, including white silk discs for comparisons. Without any irradiation, both white silk and mKate2 silk discs were kept in dark under the same conditions as two different control groups. After green light activation, each silk disc was transferred to a phosphate buffered saline (PBS; pH 7.4) solution (1 mL) and *E. coli* cells were eluted by shaking incubation for 60 minutes. To achieve a reasonable number of surviving cells for counting the colonies, the eluted cells were diluted up to 1000-fold, were plated on the LB agar, and were incubated overnight at 37 °C. CFU from the mKate2 silk disc irradiated under weak green light for 60 minutes was clearly lower than that of the mKate2 silk disc in dark (Figure 2b). Because our biological experiments were carried four different groups, we conducted ANOVA and multiple comparisons tests. In particular, Duncan multiple comparison (two-sided) tests set a 5% level of significance for all pairs of means (six possible comparisons). We performed the statistical analyses using Stata 14.2 (College Station, TX, USA).

*Detection of superoxide ($O_2^{\bullet-}$) and singlet oxygen ($^1O_2$) using fluorescent radical probes*: As free radical probes of $O_2^{\bullet-}$ and $^1O_2$, we used TEMPO-9-ac and ABDA, respectively. In the original state of TEMPO-9-ac, acridine is quenched in the presence of nitroxide moiety. $O_2^{\bullet-}$ coverts nitroxide to the corresponding piperidine, which eliminates the quenching of the blue fluorophore. Thus, blue fluorescent emission from acridine appears under ultra-violet light excitation ($\lambda_{ex} \approx 360$ nm and $\lambda_{em} \approx 440$ nm).[9] The original state of ABDA emits fluorescence under ultra-violet light excitation ($\lambda_{ex} \approx 380$ nm and $\lambda_{em} \approx 431$ nm).[23] After ABDA reacting with $^1O_2$, it is converted to an endoperoxide form that leads to a decrease in the fluorescent intensity. In this study, TEMPO-9-ac and ABDA were initially dissolved in dimethyl sulfoxide (DMSO) and were diluted in PBS, respectively, resulting in each solution containing TEMPO-9-ac (20 μM) or ABDA (20 μM). In each measurement, 12 silk discs (diameter = 5 mm and total weight = 0.06 g) or regenerated silk films were immersed in a TEMPO-9-ac or ABDA solution with stirring of 400 rpm. Because water-soluble molecules are easily smeared inside silk fibers, the adsorption-desorption equilibrium was achieved prior to green light activation; the silk discs were kept in the solution with stirring of 400 rpm in dark for two hours at least. Turn-on fluorescent signals of TEMPO-9-ac solutions and turn-off fluorescent signals of ABDA solutions were spectrofluorimetrically monitored using a spectrometer. Turn-on fluorescence (i.e. TEMPO-9-ac) from the mKate2 silk discs was also imaged and was measured using a custom-build mesoscopic (between microscopic and macroscopic) imaging setup (Figure S7, Supporting Information).[14e,34]

*Detection of superoxide ($O_2^{\bullet-}$) and singlet oxygen ($^1O_2$) using scavengers*: By detecting reduced photobleaching of mKate2 silk in the presence of $O_2^{\bullet-}$ and $^1O_2$ scavengers (ROS contributes to photobleaching[24]), we further validated Type I and Type II photosensitization reactions. In particular, we took advantage of TEMPO-9-ac and ABDA as fluorogenic scavengers (i.e. physical quenchers) of $O_2^{\bullet-}$ and $^1O_2$, respectively (Figure S8, Supporting Information).[25] Under green light irradiation ($\lambda_{ex}$ = 532 nm and optical intensity ≈ 0.2 mW mm$^{-2}$), the photobleaching effect of mKate2 silk was reduced in the presence of TEMPO-9-ac (20 μM); the fluorescent emission was relatively maintained over the irradiation time in the presence of the physical scavenger of $O_2^{\bullet-}$. Similarly, $^1O_2$ generation was detected by the maintained fluorescent intensity of mKate2 silk in the presence of ABDA (20 μM). In addition, we confirmed reduced photobleaching rates of mKate2 silk using NBT (200 μM) and NaN$_3$ (200 mM), which are often used as a scavenger of $O_2^{\bullet-}$ and $^1O_2$, respectively (Figure S9, Supporting Information).[26]

*Regeneration of mKate2 silk*: To use the polymeric nature of silk, we regenerated mKate2 silk by extracting mKate2 silk fibroin from silk cocoons. mKate2 silk cocoons were cut to pieces with sizes less than 5 mm and were heated for four hours at ~ 45 °C in a aqueous solution of NaHCO$_3$ (50 mM) with alcalase (1.5 ml L$^{-1}$) with stirring of 400 rpm. Subsequently, the silk fibers were washed with de-ionized water (~ 35 °C) several times and were dried in dark under the ambient conditions for 24 hours. We also note that conventional fibroin dissolution methods are not ideal for mKate2 silk, because these require chemical-based solution treatments at temperature of 60 °C.[13c-e] In our case, the silk fibers were completely dissolved in a lithium bromide (LiBr, 9.5 M) solution with DTT (1 mM) at 45 °C. The dissolved solution was filtered through a miracloth and was dialyzed with de-ionized water for two days to remove the remaining salt. The final concentration of mKate2 silk fibroin in the solution was ∼ 4 – 5% (w v$^{-1}$). When we followed the same method for wild-type white silk under the identical conditions, a similar final concentration of silk fibroin was obtained. The solution was stored at 4 °C in dark before use. The fabrication process of the mKate2 silk solution was carried out in dark environment to minimize photobleaching of mKate2 in silk by the room light. To form silk films, the solution was dried at 30 °C for 12 hours in an oven.

**Supporting Information**

Supporting Information is available from the Wiley Online Library or from the author.

**Acknowledgements**

This work was supported by Cooperative Research Program for Agriculture Science & Technology Development (PJ012089) from Rural Development Administration, South Korea and US Air Force Office of Scientific Research (FA2386-16-1-4114), USA.

**Conflicts of interest**

The authors declare no conflict of interest.

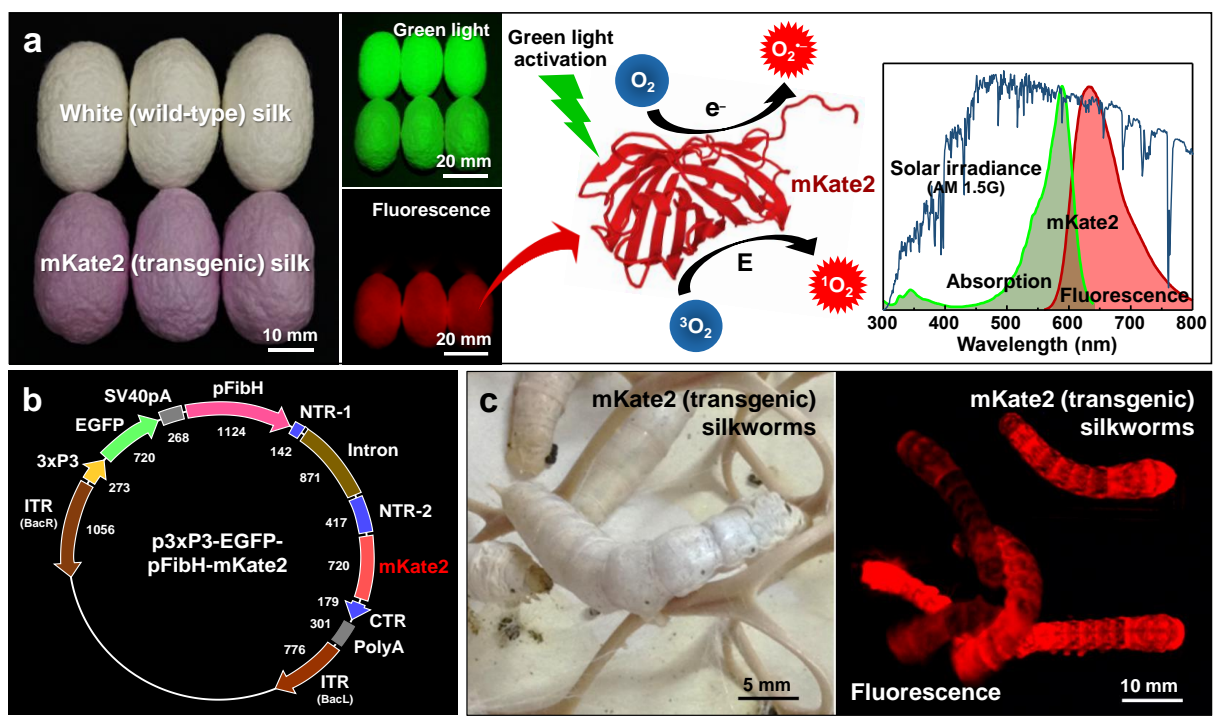

**Figure 1. Genetically encoded hybridization of far-red fluorescent protein (mKate2 and PDB ID: 3BXB) and silk for plasmonic photocatalysis-like photosensitization**. (**a**) Schematic illustration of reactive oxygen species (ROS)-generating mKate2 (transgenic) silk under green light activation. Superoxide ($O_2^{\bullet-}$) and singlet oxygen ($^1O_2$) are generated by mechanisms of electron ($e^-$) transfer and energy (E) transfer, respectively. Photographs of white (wild-type) and mKate2 (transgenic) silk cocoons and fluorescent image of mKate2 silk cocoons. Green light belongs to the peak wavelength range of the solar spectrum. (**b**) Construction of transfer vector p3xP3-EGFP-pFibH-mKate2 for mKate2 silkworm transgenesis. (**c**) Photograph and fluorescent image of mKate2 (transgenic) silkworms.

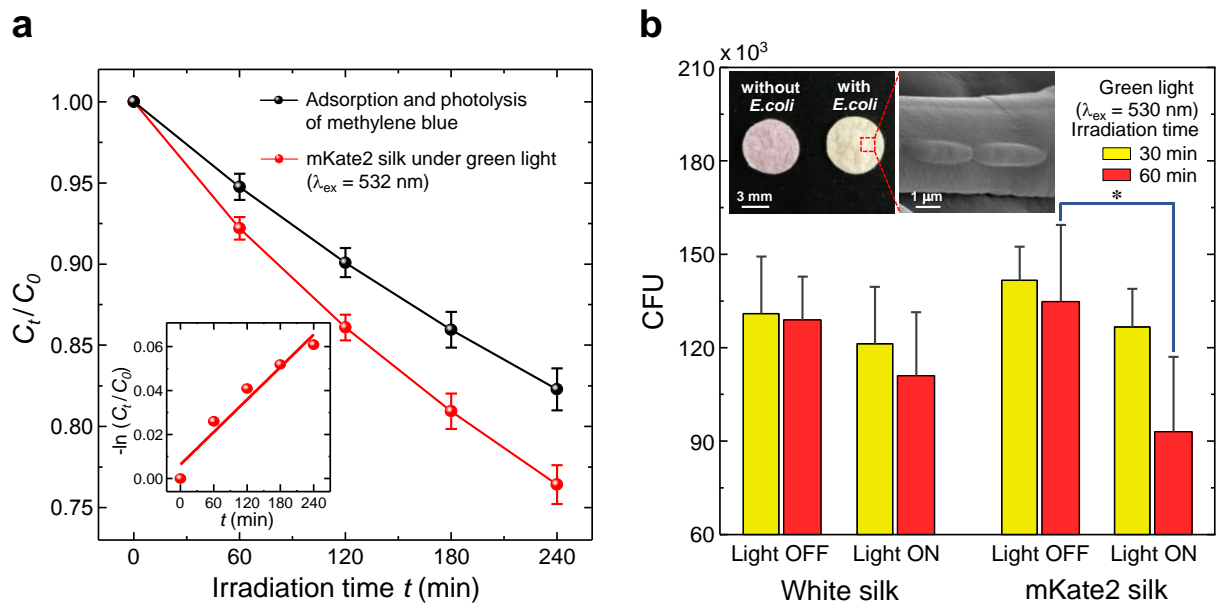

**Figure 2. Photocatalytic activity of mKate2 silk for degrading methylene blue and inactivating bacteria under green light activation at ambient temperature**. (**a**) Photodegradation of methylene blue in aqueous solutions by mKate2 silk under green laser irradiation. (Inset) Kinetic plot for methylene blue photodegradation by mKate2 silk after factoring out both adsorption and photolysis of methylene blue. $C_t/C_0$ is a relative concentration of methylene blue in an aqueous solution, where $C_0$ and $C_t$ are the concentrations of methylene blue before and after green light irradiation, respectively. The error bars are standard deviations. (**b**) Colony-forming units (CFU) of live *E. coli* (DH5α) are counted in white silk and mKate2 silk discs with and without weak green LED light activation for different irradiation periods of 30 and 60 minutes. (Inset) Representative photograph of mKate2 silk discs with and without *E. coli* and scanning electron microscopy (SEM) image of mKate2 silk attached with *E. coli* before light irradiation. Statistically significant reduction in the survival of *E. coli* occurs between 60-minuite irradiated (Light ON) and unirradiated (Light OFF) mKate2 silk (multiple comparison *p*-value = 0.031). The error bars represent standard deviations from three assays with four replicates (12 samples) in each group.

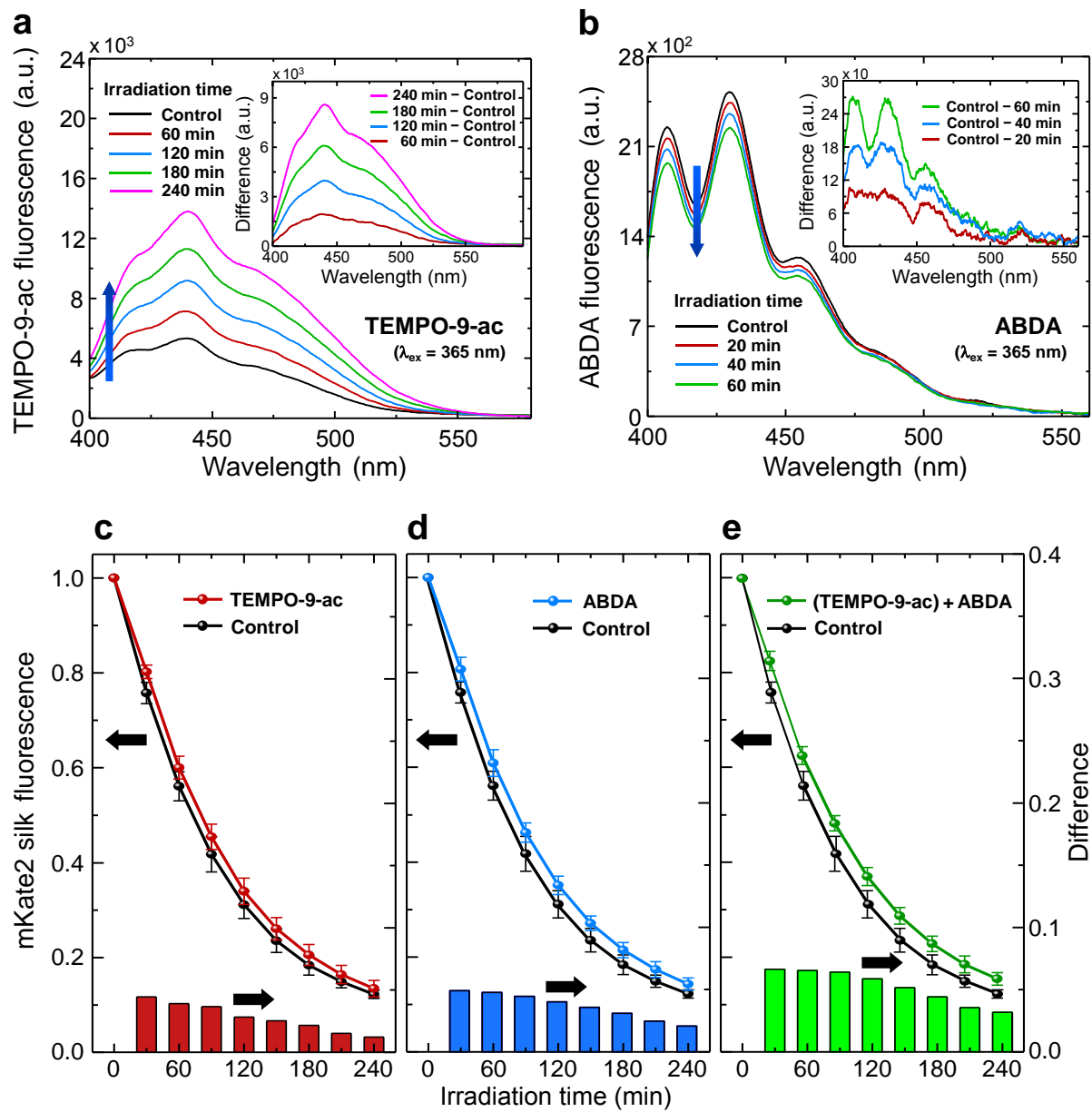

**Figure 3. Turn-on/off fluorescence detection and fluorogenic scavenger detection of ROS generated by mKate2 silk upon green light activation**. (**a&b**) Fluorescent emission signals of radical probes are recorded from solutions containing mKate2 silk discs. (**a**) $O_2^{\bullet-}$ mediated by Type I photosensitization reaction, captured by turn-on fluorescent signals of TEMPO-9-ac. (**b**) $^1O_2$ mediated by Type II photosensitization reaction, detected by reduction of the original ABDA fluorescence. (Insets) Difference in fluorescent spectra with respect to controls before green light activation. (**c-e**) Reduction in photobleaching of mKate2 silk discs is quantified by the normalized fluorescent intensity of mKate2 silk in the presence of fluorogenic scavengers of TEMPO-9-ac for $O_2^{\bullet-}$ (**c**), ABDA for $^1O_2$ (**d**), and a mixture of TEMPO-9-ac and ABDA (**e**). As a control, the normalized fluorescent intensity of mKate2 silk without the fluorogenic scavengers is plotted in black. The error bars are standard deviations. (Bottom insets) Differences in fluorescent intensity with respect to the control.

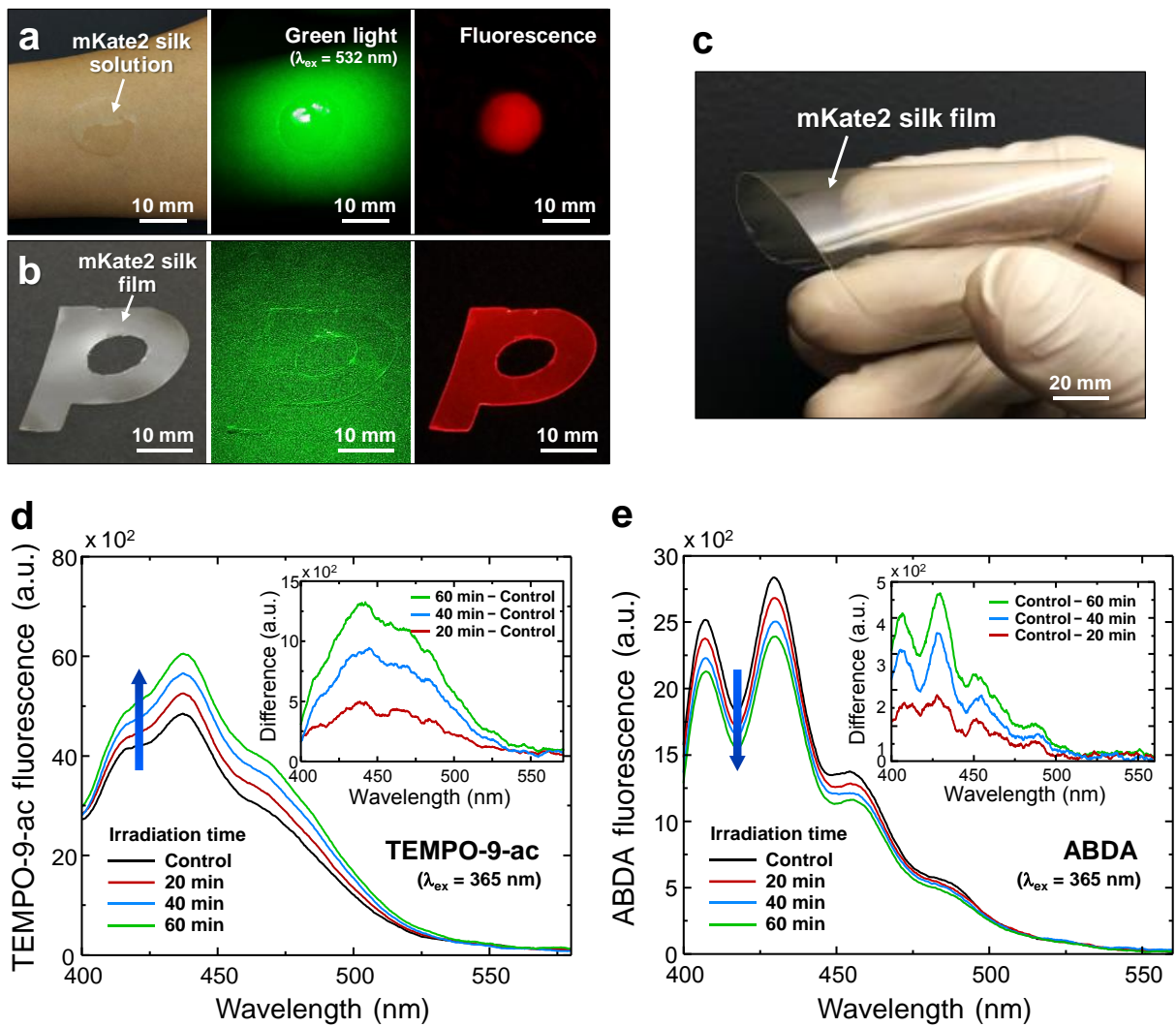

**Figure 4. Regenerated mKate2 silk and detection of ROS generation upon green light activation**. (**a**&**b**) Photographs and fluorescent images of mKate2 silk solution and film. (**c**) Photograph of large-area flexible mKate2 silk film with a diameter of 120 mm. (**d**&**e**) For regenerated mKate2 silk films, fluorescent emission signals of radical probes of TEMPO-9-ac for $O_2^{\bullet-}$ (**d**) and ABDA for $^1O_2$ (**e**). (Insets) Differences in fluorescent spectra with respect to controls before green light activation.

**Materials**

For silkworm transgenesis for producing mKate2 silk, we used *Bombyx mori* bivoltine strain, Keumokjam (F1 hybrid between the Japanese parental line *Jam 125* and the Chinese parental line *Jam 140*) from the National Institute of Agricultural Sciences (Wanju, Republic of Korea). DNA-injected eggs were kept at 25 °C in moist Petri dishes. The hatched larvae (i.e. silkworms) were reared in groups and fed with mulberry leaves under standard conditions (e.g. 25 ± 2 °C and 80 ± 10% relative humidity). For wild-type white silk, *Bombyx mori* (Baekokjam, *Jam 123 × Jam 124*) was used.

We used the following chemicals as received: alcalase enzyme, dialysis tube (pore size 12,000 Da MWCO), dimethyl sulfoxide (DMSO; $(CH_3)_2SO$, 99%), dithiothreitol (DTT; $C_4H_{10}O_2S_2$, ≥ 98%), lithium bromide (LiBr, ≥ 99%), nitro blue tetrazolium chloride (NBT; $C_{40}H_{30}Cl_2N_{10}O_6$, ≥ 98%), methylene blue ($C_{16}H_{18}ClN_3S$, 0.05 wt.% in $H_2O$), miracloth (pore size 22 – 25 μm), phosphate buffered saline (PBS; pH 7.4), sodium azide ($NaN_3$, ≥ 99.5%), sodium carbonate ($Na_2CO_3$, ≥ 99%), Triton X100, and 9,10-anthracenediyl-bis(methylene)dimalonic acid (ABDA; $C_{22}H_{18}O_8$, ≥ 90%) were purchased from Sigma-Aldrich Co. (Milwaukee, USA). 4-[(9-acridinecarbonyl)amino]-2,2,6,6-tetramethylpiperidin-1-oxyl (TEMPO-9-ac; $C_{23}H_{26}N_3O_2$, 95%) was purchased from Synchem UG & Co. KG (Altenburg, Germany). De-ionized (DI) water (Milli-Q® system) was used. All experiments were performed under the ambient conditions (22 ± 2 °C and 40 ± 10% relative humidity).

**Construction of plasmid vector DNA for silk transgenesis**

We constructed the transition vector p3xP3-EGFP-pFibH-mKate2 as the *piggyBac*-derived vector and injected the vector DNA with a helper vector into pre-blastoderm embryos, as shown in the construction sequence map (Figures S1 and S2). To obtain the fibroin pormoter, the DNA fragment (GenBank Accession No. AF226688, nucleotides 61312–63870) including pFibH promoter domain (1124 bp), N-terminal region 1 (NTR-1, 142 bp), first intron (871 bp), and N-terminal region 2 (NTR-2, 417 bp) was amplified by polymerase chain reaction (PCR) using the genomic DNA from *Bombyx mori* and specific primers (pFibHN-F: 5′-GGCGCGCCGTGCGTGATCAGGAAAAAT-3′ and pFibHN-R: 5′-TGCACCGACTGCAGCACTAGTGCTGAA-3′), followed by treatments with restriction enzymes of *Asc*I/*Not*I. The resultant DNA fragment was cloned into pGEM-T Easy Vector System (Promega, Co), named as pGEMT-pFibH-NTR. The DNA fragment (GenBank Accession No. AF226688, nucleotides 79021–79500) including C-terminal region (179 bp, CTR) and poly(A) signal region (301 bp) of the heavy chain was amplified by PCR using genomic DNA from *Bombyx mori* and specific primers (pFibHC-F: 5′-CCTGCAGGAAGTCGACAGCGTCAGTTACGGAGCTGGCAGGGGA-3′ and pFibHC-R: 5′- GGCCGGCC TATAGTATTCTTAGTTGAGAAGGCATA -3′) and then the resultant DNA fragment was cloned into pGEM-T Easy Vector System with the restriction enzymes of *Sal*I/*Sbf*I/*Fse*I, named as pGEMT-CTR. These two fragments were cloned with pBluescriptII SK(-) (Stratagene, CA) digested with *Apa*I/*Sal*I, creating pFibHNC-null. The mKate2 gene was synthesized from BIONEER Co., and then it was cloned into pGEM-T Easy Vector System pGEMT-mKate2 (720 bp). N- and C-terminal had the *Not*I and *Sbf*I restriction sites, respectively. The mKate2 cDNA was digested with *Not*I/*Sbf*I and was subcloned into a pFibHNC-null digested with *Not*I/*Sbf*I, resulting in pFibHNC-mKate2. The pFibHNC-mKate2 vector was digested with *Asc*I/*Fse*I and was subcloned into pBac-3xP3-EGFP. The resultant vector was named as p3xP3-EGFP-FibH-mKate2.

**Light sources for green light irradiation**

For optical excitation of mKate2 silk, we used two green light sources with different optical intensity: i) A diode-pumped solid-state laser coupled with a 10× zoom Galilean beam expander was used ($\lambda$ = 532 nm and optical intensity ≈ 0.2 mW mm$^{-2}$ on the sample surface). ii) As an easily accessible common light source, a green light-emitting diode (LED) was used ($\lambda$ = 530 nm with a FWHM of 30 nm and optical intensity ≈ 0.02 mW mm$^{-2}$ on the sample surface).

**Scanning electron microscopy and fluorescence confocal microscopy**

We imaged the surface morphologies of silk cocoons using a scanning electron microscopy (SEM) system (FEI Quanta 3D FEG; Oregon, USA) at 10 keV. Exploiting the fluorescent emission of mKate2 silk, we performed confocal imaging using an Olympus Fluoview FV1000 confocal laser scanning system adapted to an Olympus IX81 inverted microscope with a 20× UPlanSApo water immersion objective (Olympus, Tokyo, Japan). In this system, a green laser excitation source ($\lambda_{ex}$ = 543 nm) was used with a detection bandpass of 600 – 700 nm. The typical configuration of confocal microscopy can be summarized as follows: confocal aperture size = 50 μm (i.e. ~ 0.5 airy unit), NA = 0.4, and scan speed (pixel dwell time) = 10 μs pixel$^{-1}$. 43 image slices were stacked with a slice thickness of 5 μm along the z-axis, covering an area up to ~ 1270 μm × 1270 μm. The three-dimensional (3D) stacked image was visualized using Imaris 5.0.

**Measurements of mechanical properties**

To evaluate the basic mechanical properties of mKate2 silk fibers, we used a universal electromechanical test machine 100P/Q (TestResources Inc.) with a gauge length of 10 mm and an extension rate of 1 mm min$^{-1}$ under ambient conditions. For both white silk and mKate2 silk fibers, we tested at least 10 randomly selected single fibers from three different cocoons. As shown in Figure S10a, mKate2 silk fibers exhibited no considerable change in the mechanical properties, such as the maximum strain, the maximum stress, and the Young's modulus (*p*-value = 0.4). Thus, mKate2 silk fibers can be treated as conventional silk fibers that can be woven or constructed into large-area and continuous fabrics (e.g. knitted dress and suit) using the textile technologies (Figure S10b).

**Photodegradation of methylene blue by regenerated mKate2 silk films**

We validated the photodegradation of methylene blue by mKate2 silk films under the green light activation ($\lambda_{ex}$ = 532 nm and optical intensity ≈ 0.2 mW mm$^{-2}$) (Figure S12). For each elapsed irradiation time, a relative concentration $C_t/C_0$ of methylene blue was calculated using the absorption spectrum peak values $C_t$ at $\lambda$ = 668 nm normalized by the absorption value $C_0$ before light irradiation. We estimated the reaction kinetics, following the apparent pseudo-first-order rate equation of Langmuir-Hinshelwood kinetics: $\ln(C_t/C_0) = -k_{app}t$, where $k_{app}$ is the rate constant (min$^{-1}$) and *t* is the irradiation time. After factoring out both adsorption and photolysis of methylene blue, the mKate2 silk films also showed the photocatalytic activity, resulting in the rate constant $k_{app}$ value of 1.12×10$^{-3}$ min$^{-1}$ (Inset of Figure S12a).

**Table S1**

Multiple comparison tests of white silk and mKate2 silk with and without weak green LED light activation (irradiation time = 30 minutes)

| Colony forming unit (CFU) | Mean difference | t | p-value | 95% CI | |
|---|---|---|---|---|---|
| White silk + Light OFF[a]  vs. White silk + Light ON[b] | -9,683 | -0.77 | 0.472 | -3,6252 | 16,885 |
| mKate2 silk + Light OFF  vs. White silk + Light ON | 10,692 | 0.85 | 0.398 | -14,573 | 35,957 |
| mKate2 silk + Light ON vs. White silk + Light ON | -4,317 | -0.34 | 0.732 | -29,582 | 20,948 |
| mKate2 silk + Light OFF  vs. White silk + Light ON | 20,375 | 1.63 | 0.145 | -7,049 | 47,799 |
| mKate2 silk + Light ON vs. White silk + Light ON | 5,367 | 0.43 | 0.671 | -19,898 | 30,632 |
| mKate2 silk + Light OFF  vs. mKate2 silk + Light ON | -15,008 | -1.2 | 0.266 | -41,577 | 11,560 |

[a]without green LED light irradiation.
[b]with green LED light irradiation.

**Table S2**

Multiple comparison tests of white silk and mKate2 silk with and without weak green LED light activation (irradiation time = 60 minutes)

| Colony forming unit (CFU) | Mean difference | $t$ | $p$-value | 95% CI | |
|---|---|---|---|---|---|
| White silk + Light OFF[a]  vs. White silk + Light ON[b] | -17,967 | -1.04 | 0.305 | -52,859 | 16,926 |
| mKate2 silk + Light OFF vs. White silk + Light ON | 5,884 | 0.34 | 0.736 | -29,009 | 40,776 |
| mKate2 silk + Light ON vs. White silk + Light ON | -35,850 | -2.07 | 0.055 | -72,543 | 843 |
| mKate2 silk + Light OFF vs. White silk + Light ON | 23,850 | 1.38 | 0.201 | -12,843 | 60,543 |
| mKate2 silk + Light ON vs. White silk + Light ON | -17,883 | -1.03 | 0.307 | -52,776 | 17,009 |
| mKate2 silk + Light OFF vs. mKate2 silk + Light ON | -41,733 | -2.41 | *0.031 | -79,607 | -3,860 |

[a]without green LED light irradiation.
[b]with green LED light irradiation.

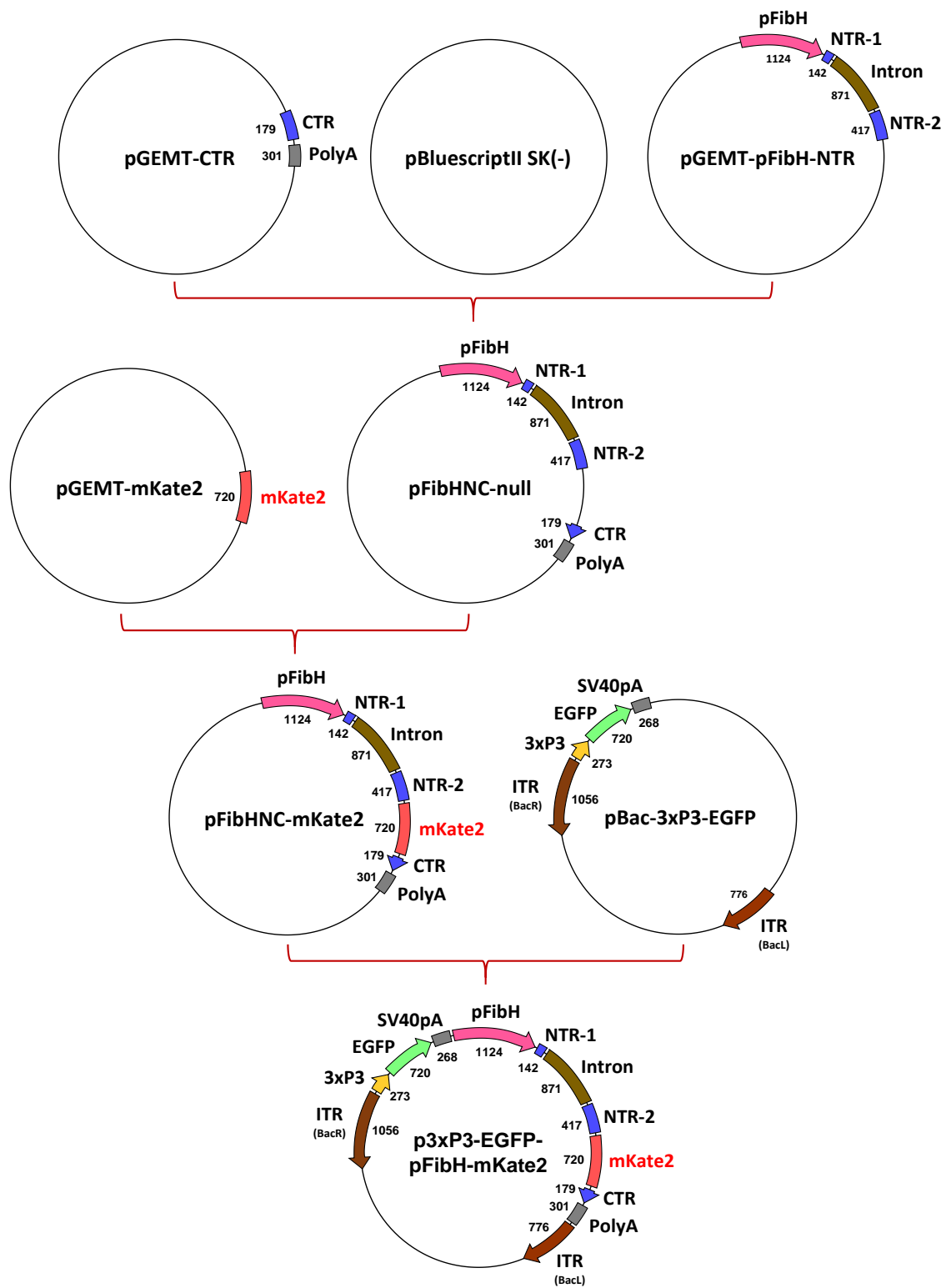

**Figure S1. Construction sequence map of transfer vector p3xP3-EGFP-pFibH-mKate2.** The nucleotide sequences of pFibH-NTR and CTR are derived from Genebank Accession No. AF226688. pFibH: fibroin heavy chain promoter domain (1124 bp), NTR-1: N-terminal region 1 (142 bp), intron: first intron (871 bp), NTR-2: N-terminal region 2 (417 bp), CTR: C-terminal region (179 bp), PolyA: poly(A) signal region (301 bp), EGFP: enhanced green fluorescent protein gene, mKate2: monomeric far-red fluorescent protein, ITR (BacR, BacL): inverted repeat sequences of piggyBac arms, 3xP3: 3xP3 promoter, and SV40: SV40 polyadenylation signal sequence.

**a**

| No | Location | Mr(expt) | Mr(calc) | Sequence |
|---|---|---|---|---|
| 1 | 49-63 | 1590.8493 | 1589.8148 | DASGAVIEEQITTKK |
| 2 | 70-79 | 2442.8413 | 2441.1519 | NHGILGKNEK |
| 3 | 83-104 | 2442.8413 | 2441.1519 | TFVITTDSDGNESIVEEDVLMK |
| 4 | 159-171 | 1637.3093 | 1637.7462 | MVSELIKENMHMK |
| 5 | 172-184 | 1589.3663 | 1589.7296 | LYMEGTVNNHHFK |
| 6 | 185-201 | 1946.4763 | 1945.8145 | CTSEGEGKPYEGTQTMR |
| 7 | 204-226 | 2433.9883 | 2433.1925 | AVEGGPLPFAFDILATSFMYGSK |
| 8 | 252-279 | 3062.0203 | 3060.4597 | VTTYEDGGVLTATQDTSLQDGCLIYNVK |
| 9 | 280-294 | 1660.4443 | 1660.8242 | IRGVNFPSNGPVMQK |
| 10 | 295-316 | 2350.9313 | 2351440 | KTLGWEASTETLYPADGGLEGR |
| 11 | 348-357 | 1270.9383 | 1270.6128 | MPGVYYVDRR |
| 12 | 363-379 | 1973.5673 | 1973.9330 | EADKETYVEQHEVAVAR |
| 13 | 380-386 | 881.6313 | 881.3953 | YCDLPSK |
| 14 | 387-403 | 1826.5823 | 1827.8863 | LGHRPQQVDSVSYGAGR |
| 15 | 404-422 | 1687.4983 | 1685.7604 | GYGQGAGSAASSVSSASSR |
| 16 | 423-429 | 945.4853 | 945.4304 | SYDYSRR |
| 17 | 433-439 | 843.3973 | 843.4385 | KNCGIPR |

**b**

MRVKTFVILC CALQYVAYTN ANINDFDEDY FGSDVTVQSS NTTDEIIRDA SGAVIEEQIT TKKMQRKNKN
HGILGKNEKM IKTFVITTDS DGNESIVEED VLMKTLSDGT VAQSYVAADA GAYSQSGPYV SNSGYSTHQG
YTSDFSTSAA VGAGSSGRMV SELIKENMHM KLYMEGTVNN HHFKCTSEGE GKPYEGTQTM RIKAVEGGPL
PFAFDILATS FMYGSKTFIN HTQGIPDFFK QSFPEGFTWE RVTTYEDGGV LTATQDTSLQ DGCLIYNVKI
RGVNFPSNGP VMQKKTLGWE ASTETLYPAD GGLEGRADMA LKLVGGGHLI CNLKTTYRSK KPAKNLKMPG
VYYVDRRLER IKEADKETYV EQHEVAVARY CDLPSKLGHR PQQVDSVSYG AGRGYGQGAG SAASSVSSAS
SRSYDYSRRN VRKNCGIPRR QLVVKFRALP CVNCN

▪ N-terminal domain of the fibroin H-chain    ▪ C-terminal domain of the fibroin H-chain    ▪ mKate2

**Figure S2. Mass spectrometric analyses.** (**a**) Peptides from mKate2. (**b**) Sequence alignment of mKate2/Fibroin H-chain fusion recombinant protein amino acid. The mass density of mKate2/Fibroin H-chain fusion recombinant protein in mKate2 silk is estimated to be ~ 12.6 %.

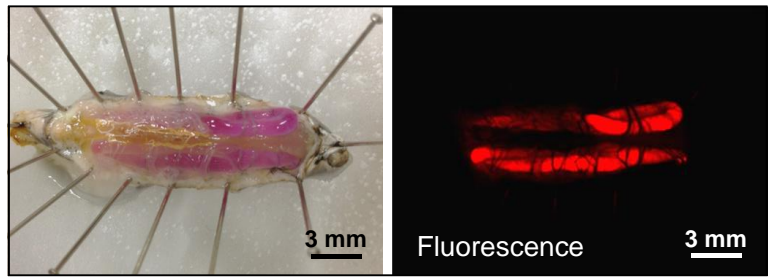

**Figure S3**. **mKate2 fluorescence in the silk gland of the transgenic line**. Photograph (left) and fluorescent image (right) of the silk gland for the transgenic mKate2 silkworm larvae at the 3$^{rd}$ day of the 5$^{th}$ instar.

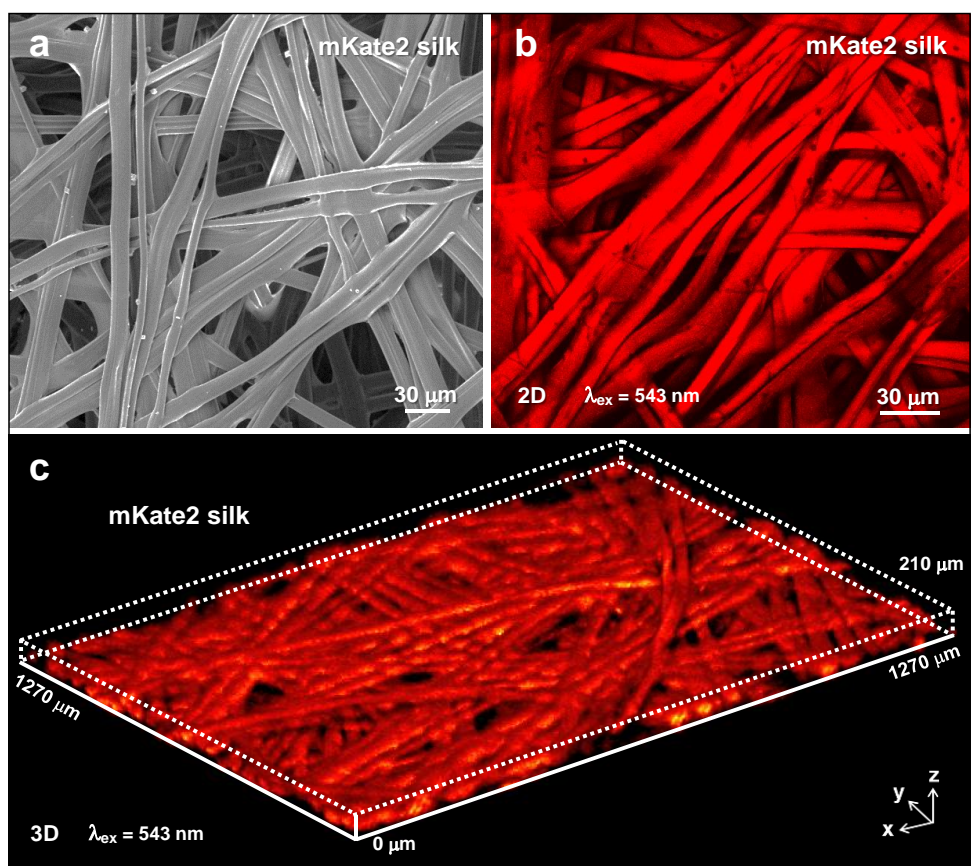

**Figure S4**. **Microscopic images of mKate2 silk.** (**a**) SEM image of mKate2 silk fibers. (**b&c**) Confocal fluorescence microscopy images of mKate2 silk fibers under green light excitation.

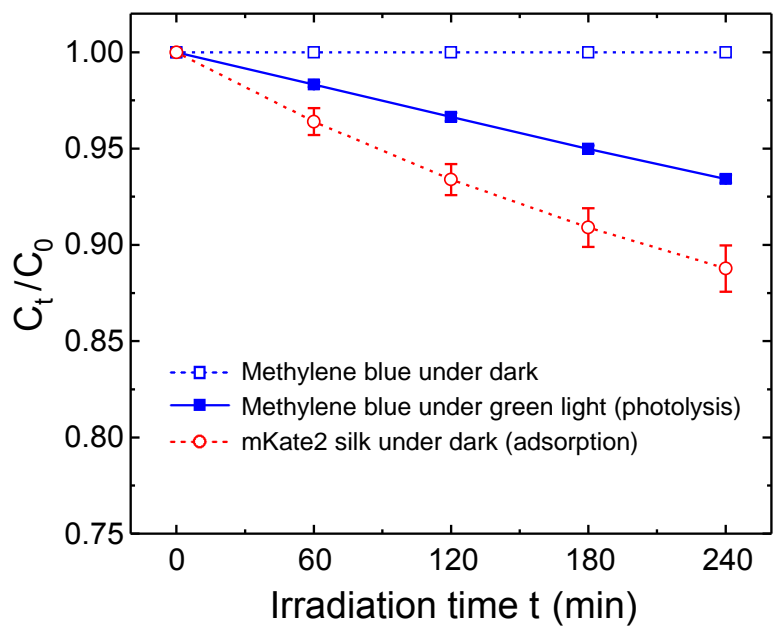

**Figure S5. Confounding factors in photodegradation of methylene blue by mKate2 silk.** The adsorption of methylene blue to mKate2 silk and the photolysis of methylene blue under green light irradiation are separately measured. The error bars are standard deviations.

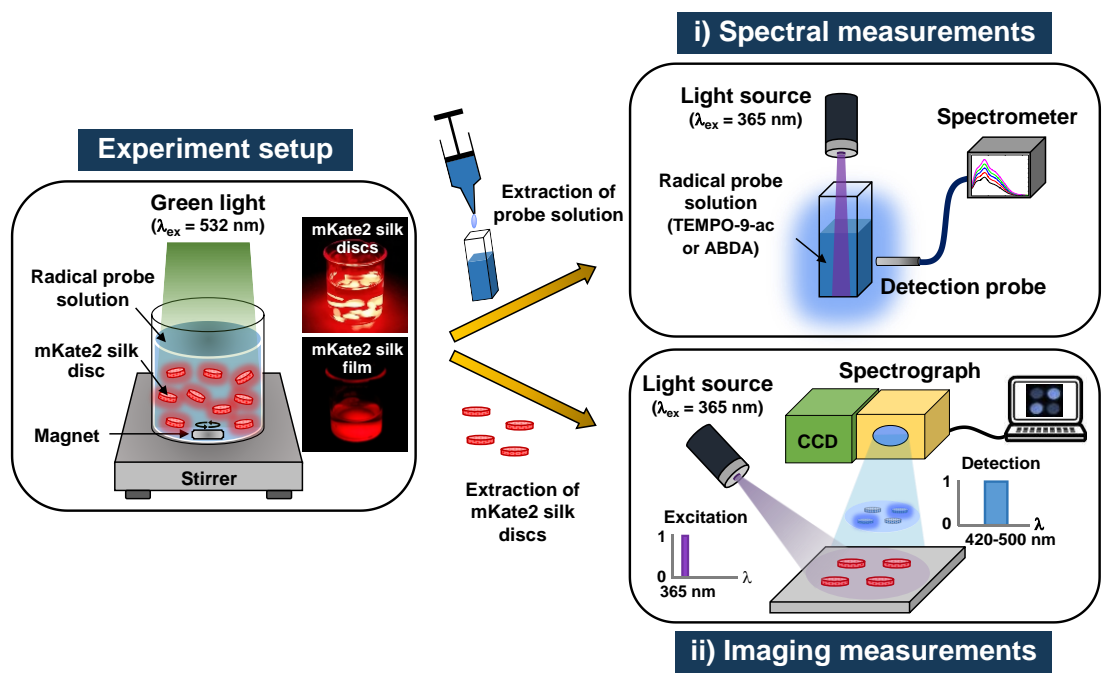

**Figure S6**. **Two detection scenarios for assessing ROS generated by mKate2 silk upon green light activation**. i) Spectral measurements: Fluorescent signals of radical probes under excitation of $\lambda_{ex}$ = 365 nm are detected from the solution including mKate2 silk discs. ii) Imaging measurements: Fluorescent radical probes are permeated into mKate2 silk discs. Specimens are arranged within the field of view of the mesoscopic imaging setup, in which the excitation ($\lambda_{ex}$ = 365 nm) and emission filters ($\lambda_{em}$ = 420 – 500 nm) are used as illustrated.

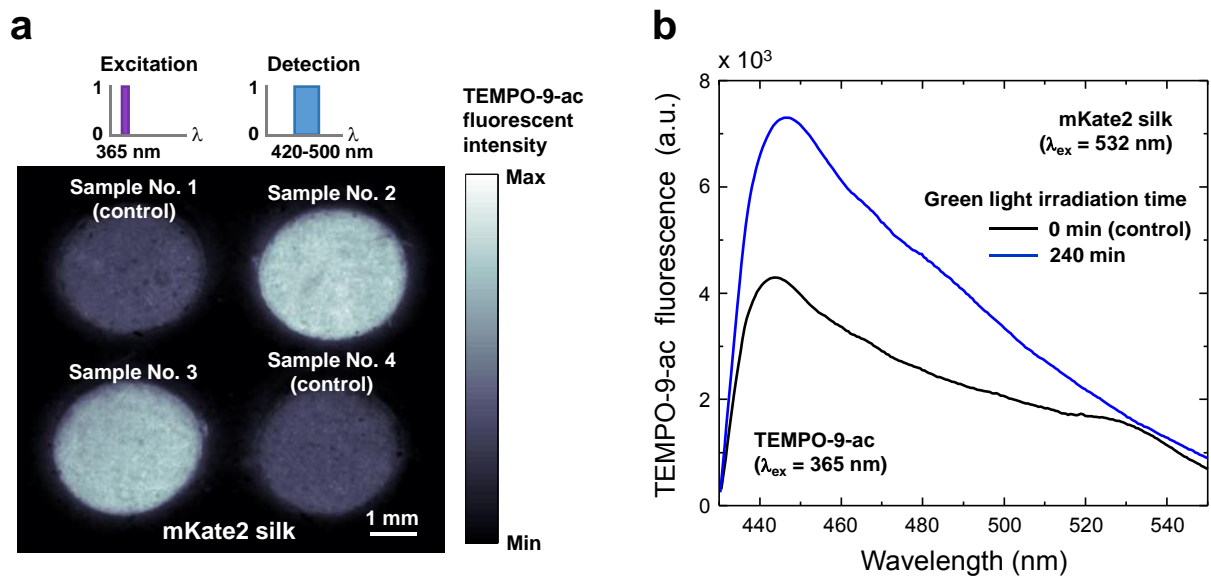

**Figure S7**. **Turn-on fluorescent signals of TEMPO-9-ac in mKate2 silk.** (**a**&**b**) Fluorescent images (**a**) and spectra (**b**) of TEMPO-9-ac ($\lambda_{ex}$ = 365 nm) in mKate2 silk discs without green light irradiation (controls) and with green light irradiation for 240 minutes.

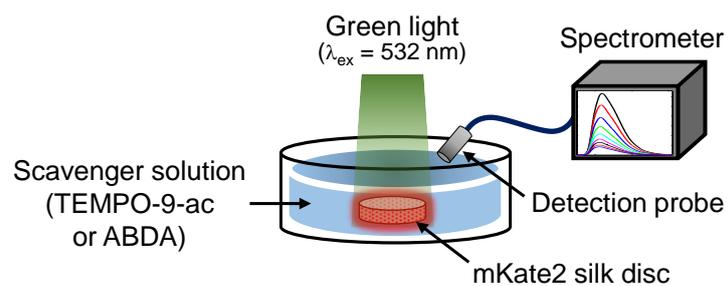

**Figure S8**. **Photobleaching of mKate2 fluorescence.** TEMPO-9-ac and ABDA, which are used as physical scavengers of phototoxic ROS generated by mKate2 silk, slow down photobleaching of mKate2 in silk.

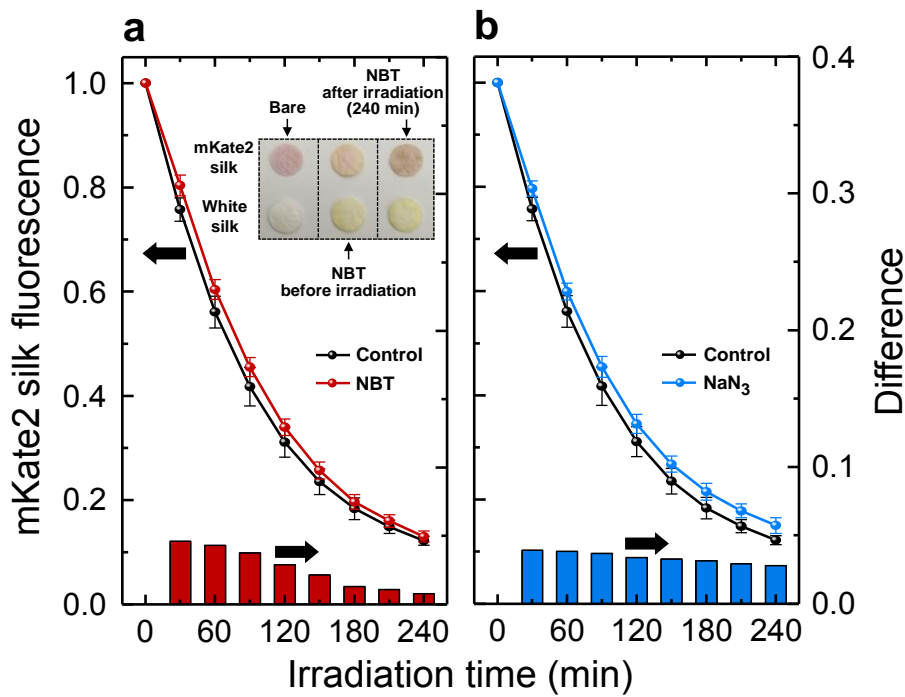

**Figure S9**. **Scavenger detection of ROS generated by mKate2 silk upon green light activation**. **(a&b)** Normalized fluorescent intensity of mKate2 silk with and without scavengers of NBT for $O_2^{\bullet-}$ (**a**) and NaN$_3$ for $^1O_2$ (**b**), respectively. As a control, the normalized fluorescent intensity of mKate2 silk without the scavengers is plotted in black. The error bars are standard deviations. (Bottom insets) Differences in fluorescent intensity with respect to the control. (Top inset of **a**) Photograph of bare and NBT-treated (before and after light irradiation) white silk and mKate2 silk discs, supporting the $O_2^{\bullet-}$ generation. After 240-minute green light irradiation, there are no variations in the color (yellow) of white silk, while mKate2 silk changes to the bluish color, resulting from the formation of blue chromagen diformazan.

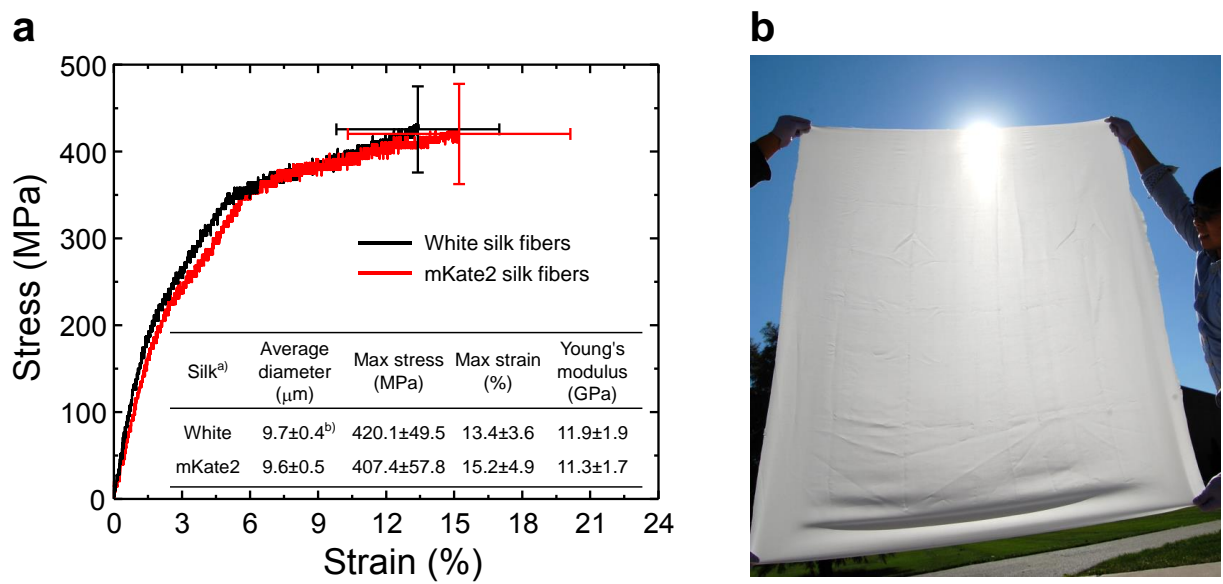

**Figure S10. Mechanical tests of mKate2 silk and scalable/continuous manufacturing of silk fabrics**. (**a**) Representative strain-stress curves of white silk and mKate2 silk fibers. The error bars are standard deviations in the elongation at break (horizontal axis) and the fracture strength (vertical axis). [a]For each silk, at least 10 randomly selected single fibers from three silk cocoons are tested for statistical analyses. [b]Mean ± standard deviation. The Young's moduli are calculated from the first linear regime of the strain-stress curve before the first bend. (**b**) Photograph of 110 cm × 140 cm silk fabric woven in the Korea Silk Research Institute (Jinju, Republic of Korea). This white silk fabric did not undergo any additional chemical treatments except for sericin removal (i.e. degumming), showing the possibility of scalable and continuous fabrication using the conventional textile infrastructures.

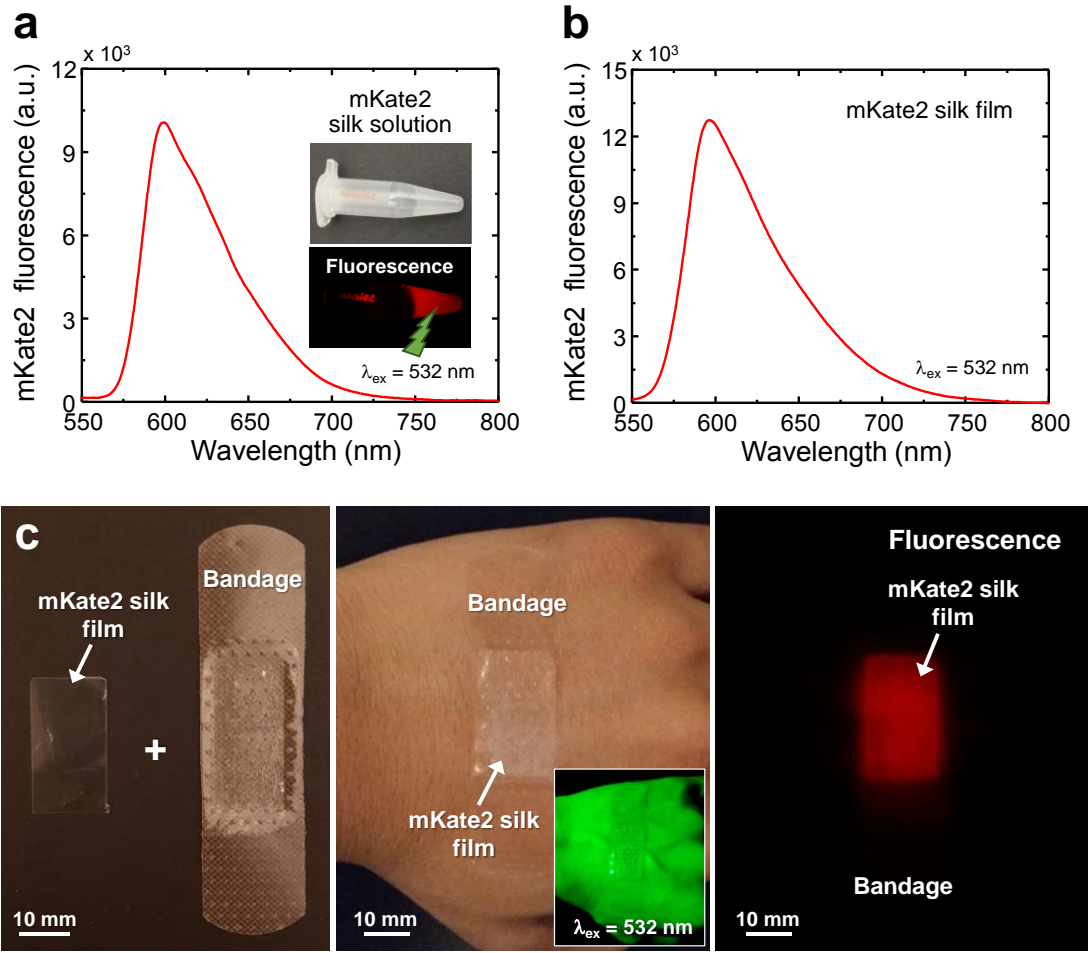

**Figure S11**. **Fluorescent spectra of regenerated mKate2 silk and representative utilization of regenerated mKate2 silk.** (**a**&**b**) Fluorescent spectra of regenerated mKate2 silk in forms of solution (**a**) and film (**b**), respectively. (Inset of **a**) Photograph and fluorescent image of mKate2 silk solutions. (**c**) A regenerated mKate2 silk film can be integrated with a bandage, potentially offering an additional functionality of controllable ROS release using a simple light source.

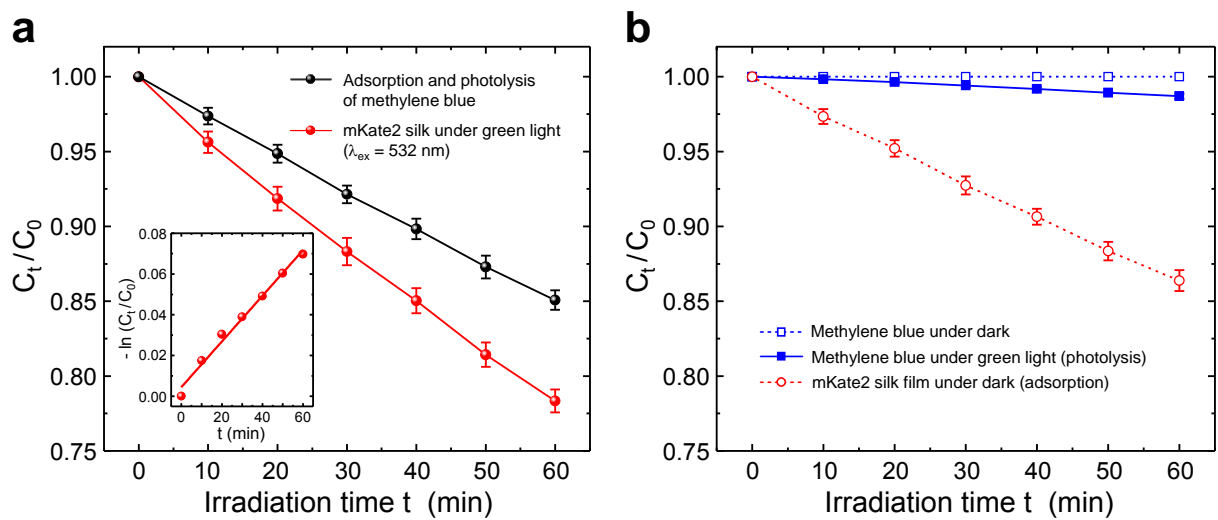

**Figure S12**. **Photocatalytic activity of regenerated mKate2 silk for degrading methylene blue under green light activation at the ambient temperature.** (**a**) Photodegradation of methylene blue upon green light activation. (Inset of **a**) Kinetic plot for methylene blue photodegradation by mKate2 silk film after factoring out both adsorption and photolysis of methylene blue ($k_{app}$ = 1.12×10$^{-3}$ min$^{-1}$). (**b**) Confounding factors in photodegradation of methylene blue by mKate2 silk film for the adsorption of methylene blue to the mKate2 silk film and the photolysis of methylene blue under the green light irradiation. The error bars are standard deviations.